\documentstyle{mn}

\title[Lensing Studies of Clusters: II Cluster Mass Distributions.]
{Gravitational Lensing of Distant Field Galaxies by\\
 Rich Clusters: II. Cluster Mass Distributions.}

\author[Smail et al.]
{Ian Smail,$\!\!$\thanks{Present address: Caltech 105-24, Pasadena CA
91125.} Richard S.\ Ellis\thanks{Present address:  Institute of
Astronomy, Madingley Road, Cambridge CB3 0HA, UK}
and Michael J.\ Fitchett\thanks{Present address:
Medical School, University of Newcastle, Framlington Place, Newcastle, UK} \\
Physics Department, University of Durham, South Road, Durham DH1 3LE, UK
\and Alastair C.\ Edge \\ Institute of
Astronomy, Madingley Road, Cambridge CB3 0HA, UK}

\date {Received 1994 --- --; in original form 1994 February 20.}

\def\UV{\hbox{$U-V$}}

\def\VI{\hbox{$V-I$}}
\def\RI{\hbox{$R-I$}}
\def\VR{\hbox{$V-R$}}
\def\Msunpyr{\hbox{$M_\odot$ yr$^{-1}$}}

\def\gs{\mathrel{\raise0.3ex\hbox{$\scriptstyle >$}\kern-0.70em %
\lower0.71ex\hbox{{$\scriptstyle \sim$}}}}
\def\ls{\mathrel{\raise0.3ex\hbox{$\scriptstyle <$}\kern-0.70em 
\lower0.71ex\hbox{{$\scriptstyle \sim$}}}}
\def\et{\hbox{\it et al.}$\,$}

\begin{document}
\label{firstpage}
\maketitle

\begin{abstract}

We construct a photometric catalogue of very faint galaxies ($I \ls
25.5$) using deep CCD images taken with the 4.2m William Herschel
telescope taken of fields centred on two distant X-ray luminous clusters:
1455+22 ($z_{cl}=0.26$) and 0016+16 ($z_{cl}=0.55$).  Using a
non-parametric procedure developed by Kaiser \& Squires (1993), we
analyse the statistical image distortions in our samples to derive two
dimensional projected mass distributions for the clusters. The mass maps
of 1455+22 and 0016+16 are presented at effective resolutions of 135
kpc and 200 kpc respectively (for $H_o$=50 kms sec$^{-1}$ Mpc$^{-1}$,
$q_o=0.5$) with a mean signal to noise per resolution element of 17 and 14.
Although the {\it absolute} normalisation of these mass maps depends on the
assumed redshift distribution of the $I\ls25.5$ field galaxies used as
probes, the maps should be reliable on a {\it relative} scale and will
trace the cluster mass regardless of whether it is baryonic or non-baryonic.
We compare our 2-D mass distributions on scales up to $\sim$1 Mpc with those
defined by the spatial distribution of colour-selected cluster members and
from deep high resolution X-ray images of the hot intracluster gas.
Despite the different cluster morphologies, one being cD-dominated and
the other not, in both cases the form of the mass distribution derived
from the lensing signal is strikingly similar to that traced by both the
cluster galaxies and the hot X-ray gas. We find some evidence
for a greater central concentration of dark matter with respect to the
galaxies. The overall similarity between the distribution of total mass and
that defined by the baryonic components presents a significant
new observational constraint on the nature of dark matter and the
evolutionary history of rich clusters.

\end{abstract}


\section{INTRODUCTION}

The nature of the dark matter in clusters of galaxies has been a central
theme in cosmological research since its existence was inferred over
sixty years ago (Zwicky 1933). Although the early evidence was based on
the virial analysis of the relative velocities of cluster member galaxies,
the discovery of X-ray emission from a hot intracluster medium provided
independent verification of the so-called `missing mass problem'
(Jones \& Forman 1984).

Until recently, most of the interest in clusters of galaxies has
focussed on determining the {\it amount} of dark matter contained.  As
clusters are the largest bound structures known, their mass/light
ratios and baryonic fractions should approach that for the cosmos as a
whole.  In fact, careful studies of selected nearby clusters (Fabian
1991, White \et 1993, Mushotzky 1993) have revealed a much higher
baryonic fraction than expected in the inflationary $\Omega_{tot}$=1
Universe whose baryonic component is constrained by primordial
nucleosynthesis arguments  (White \et 1993). This dilemma might be
resolved if it could be demonstrated that the dark matter in clusters
was less concentrated than the baryonic component.
Such arguments indicate that the observations constraining
the relative {\it distribution} of dark matter are as important
as those which estimate its total amount.

Both the classical spectroscopic and X-ray techniques are ill-suited to
tackling this problem. In the case of the radial velocities of cluster
galaxies, the one-dimensional nature of the dynamical data necessitates
assumptions about the distribution of orbits which even extensive data
(Kent \& Gunn 1982, Sharples, Ellis \& Grey 1988) has been unable to resolve
definitively. Simulations have shown (Fitchett 1988) the difficulty of
recognising substructure even with several hundred radial velocities
unless it is of a particular form (e.g.\ bimodal) and conveniently
well-separated on the sky or in velocity. Beyond a few core radii
(i.e.\ $r>$500 kpc), contamination from non-members increases
to such an extent that useful samples of cluster galaxies must be taken
from well down the galaxy luminosity function making such surveys
highly inefficient probes of the mass distribution on large scales.

Analyses of the X-ray gas distribution is less complicated by
projection effects and provides a better sampling of any spatial
structure on scales less than a few hundred kiloparsecs. The
fundamental limitation of X-ray studies to date has been the lack of
spatially resolved temperatures profiles. Detailed studies of nearby
clusters have shown no evidence for strong temperature gradients in
clusters (Hughes 1989, Eyles \et 1991) and within the central 1~Mpc the
assumption of isothermality is apparently consistent with all published
data (excluding the central cooling core of 100--200~kpc radius).  The
ASCA X-ray satellite is currently generating spatially-resolved
temperature profiles for many clusters and thus considerable progress
will be made in this area in the next few years.  However, as with the
cluster galaxies, the surface brightness of the X-ray emission falls
precipitously with radius from the cluster centre and to determine mass
distributions on the large scales required to resolve the `baryon
catastrophe' discussed by White \et (1993), very long exposures
are needed to obtain the necessary temperature data.

Thus far, mass distributions derived from the modelling of the X-ray
emission have been published for a few nearby clusters and are claimed
to indicate a dark matter component more centrally concentrated than
both the galaxies and the X-ray gas (Eyles \et 1991, Gerbal \et 1992).
One recent study (Buote \& Canizares 1992) has compared the
morphologies of the 2-D distributions of mass and galaxies in a sample
of local clusters using X-ray imaging data. Although the orientations
of the two distributions (X-ray gas and galaxy number density) are in
good agreement, they conclude that the potential traced by the X-ray
gas in the central $\sim$1 Mpc of their clusters is too round to be
generated by the observed galaxy distribution if the galaxies exactly
trace the mass.  The difference might be reconciled if the galaxies and
mass have a different radial scale lengths.  This would mean that the
X-ray analysis of Buote \& Canizares would include regions
outside the cores of the mass distribution resulting
in an inferred ellipticity for the mass in better agreement with that
traced by the galaxies.

A related issue in the quest for the nature and distribution of dark
matter in clusters is the evolutionary history of the gravitational
potential of clusters. Conventional theories that postulate a
significant non-baryonic component predict remarkably recent growth
(Frenk \et 1990). X-ray imaging (Henry \et 1992) and all-sky scanning
(Edge \et 1990) surveys have revealed evidence for evolution at
surprisingly low redshifts ($z\leq0.2$), in the sense that there are
fewer of the highest X-ray luminosity ($L_x>8\times 10^{44}$ erg
s$^{-1}$) clusters in the past.  In contrast, optical surveys find a
nearly constant co-moving space density of rich clusters to
$z_{cl}\sim0.5$ (Gunn \et 1986, Couch \et 1991). To reconcile this
discrepancy, Kaiser (1991) has proposed that the X-ray emitting gas was
heated prior to the formation of clusters and became bound to the
potential wells only when the latter had grown sufficiently deep to
contain the gas.  In any case the mechanism of hierarchical merging
that underlies the observed X-ray evolution may introduce significant
differences in the optical and X-ray properties of clusters in the
period leading up to and immediately after a cluster-cluster merger.
Thus we might expect morphological differences between the
distributions of the X-ray gas and the gravitating mass in moderate
redshift clusters.   In a search for such signatures of merger activity
we have made a detailed comparison of optical, X-ray and gravitating
mass distributions for two high X-ray luminosity clusters.

The analysis of the gravitational lensing signals, derived from the
distortion of background galaxies viewed through clusters, offers an
independent probe of the mass distribution in clusters (for recent
reviews see Soucail 1992, Blandford \& Narayan 1993). Geometrical
considerations indicate the phenomenon will be most effective in
constraining mass distributions in clusters with redshifts $z_{cl} \gs
0.2$. The classical techniques for recovering mass distributions rely
on {\it in situ} probes of the gravitational potential about which
various assumptions are made concerning their thermodynamic state. For
the lensing methods, however, the results depend on the characteristics
of sources totally unrelated to the cluster. In many cases these can be
directly measured or they can be statistically understood from large
samples.

\section{GRAVITATIONAL LENSING AS A PROBE OF CLUSTER MASS DISTRIBUTIONS}

The lensing of distant galaxies by rich clusters produces two observable
phenomena: `giant arcs' and `arclets'.

As highly elongated images of serendipitously positioned background
galaxies, the giant arcs are easily recognisable manifestations of
strong lensing by cluster cores (Grossman \& Narayan 1989). Several
have spectroscopic redshifts which when combined with detailed
modelling of their image characteristics have provided important
constraints on the total mass in the centres of rich clusters.
Employing a variety of gravitationally lensed features in Abell 370,
Kneib \et (1993) claim a bimodal mass distribution which is
morphologically similar to that delineated by the X-ray emission and
cluster red light. However, for Abell 2390, Kassiola, Kovner \&
Blandford (1992) require a mass distribution which differs from that seen in
the cluster galaxies.

Unfortunately, with at most a single giant arc per cluster, such models
are not uniquely constrained by the available data.  Using a
sample of cluster with arcs Wu \& Hammer (1993) claimed a marked
concentration of dark compared to visible mass from the mean cluster
radius where the arcs are found compared to the canonical X-ray core
radius.  However, they conclude that further information on the sources
(e.g.\ their sizes) is required to derive robust conclusions. The
apparent concentration of dark compared to visible mass would
only remain a valid conclusion if the distant sources are comparable in
intrinsic size to present-day galaxies.  The recent discovery of
multiply-imaged pairs in the cores of rich clusters (Kneib \et 1993,
Smail \et 1994) suggests at least some $B\simeq$26-28 galaxies are very
compact, as these images would otherwise appear as elongated connected
arcs (Miralda-Escud\'e \& Fort 1993).

In a detailed study of three clusters with both arcs and X-ray data
available from the literature, Babul \& Miralda-Escud\'e (1994)
conclude that the mass estimates from the arc modelling can be nearly a
factor of $\simeq$2--3 larger than those from the X-ray observations of
the inner-most regions of the cluster core.
As they discuss, this discrepancy may arise from a
number of sources, including the simplified geometry adopted for the
lensing clusters or other invalid assumptions used in the X-ray
modelling.

The rarity of the giant arcs together with uncertainties about the
intrinsic source properties (including sizes and even redshifts in many
cases) precludes reliable constraints. In very rare cases such as AC114
(Smail \et 1994), the combination of giant arcs and a multiply-imaged
pair {\it resolved} with the aid of Hubble Space Telescope can yield
tight constraints on the mass distribution but only for the inner 1-200
kpc of a single cluster. The frequency of occurrence of multiply-imaged
pairs remains unclear, although there is good cause for being
optimistic if the  sources are $B\simeq$26-28 galaxies as detailed
modelling suggests.

Arclets are a much more promising probe of the mass
distribution.  These images are generally too faint for direct
spectroscopy and are insufficiently elongated to be convincingly due to
gravitational lensing on an individual basis. However, the tiny
distortions induced by the lensing cluster form a coherent pattern
superimposed upon the intrinsic ellipticities and orientations of the
faint background population. The coherence thus overcomes the
low signal to noise of the the individual arclets.

The most basic lensing statistic is the proportion of field galaxies
aligned tangentially to a suitably-defined lens centre. In a pioneering
study Tyson \et (1990) analysed images of the central regions of
Abell~1689 ($z=0.18$) and Cl1409+52 (3C295, $z=0.46$). Using the alignment of
blue galaxies, they derived radial `mass' profiles for the clusters
which were found to resemble the profiles of cluster light. As Kaiser
\& Squires (1992) show, Tyson {\it et al.}'s statistic measures the
surface potential rather than the mass.

With deep CCD images reaching $I\simeq25$, $B\simeq27$, a very high
surface density of background sources can be attained. Techniques can
then be developed to determine the mass distribution with a resolution
that matches the best available from the X-ray and optical tracers in
the cluster. The unique advantage of this method over those discussed
above is that, providing the cluster mass distribution has a non-zero
gradient, the technique works equally well in the cluster peripheries
as in the core, since the basic signal is provided by an isotropic
population of faint field galaxies.

Of course, the lensing signal depends on the redshift distribution,
$N(z)$, of the source population, as well as on the desired mass
distribution, $M(r)$, within the cluster. At the limiting magnitudes
essential for generating a high surface density of background sources,
even with 10-m telescopes it seems unlikely that spectroscopy can
usefully constrain $N(z)$.  Another possible route to achieve
arbitrarily high surface density of spectroscopically attainable
background sources is to combine shallow images of many clusters to
study the profile of an `average' cluster.

More complex analysis combining the results from individual clusters
can be used to separate the dependency on $N(z)$. We therefore began a
deep imaging programme for 3 X-ray luminous clusters, carefully
selected to cover a range of cluster redshifts, $z_{cl}$. By imaging
each cluster to the {\it same limiting magnitude}, the lensing signals
can be analysed to test both $N(z)$ to the chosen limiting magnitude
and $M(r)$ for each of the clusters.  Paper I of this series (Smail,
Ellis \& Fitchett 1994) describes the overall approach, the
observational datasets and tests in some detail.  {\it The reader is
strongly advised to consult that article first}.

In Paper I, we derive maximum-likelihood constraints on the redshift
distribution, $N(z)$, of the faint field population imaged to
$I\leq25$. The most probable distribution is close to that expected in
the absence of pure luminosity evolution, although a tail of high
redshift sources cannot be excluded. Our analysis technique also
quantified the amount and concentration of the mass in the two lower
redshift ($z$=0.26, 0.55) clusters for which a strong lensing signal is
seen. The mass distributions were, however, only derived in a form
parametrised by the depth of the potential and its scale in the
isothermal case (i.e.\ in terms of the line of sight velocity
dispersion, $\sigma_{cl}$, and core radius, $r_c$). No true
morphological information on $M(r)$ was discussed.

In Paper I the joint dependence of the lensing signal on $N(z)$ and $M(r)$
was only partially separated. Whilst the parameters satisfying the cluster
mass distributions are consistent with those inferred from dynamical and
X-ray data, the uncertainties are too great to examine any possible
differences. Nevertheless, for the redshift distributions (which formed the
basic motivation of Paper I), significant constraints are possible by
combining the results from the three clusters.

This second paper extends the analysis begun in Paper I and examines
the {\it relative mass distributions} in the two lowest redshift
clusters in a non-parametric form.  This is made possible by the
non-parametric lens inversion technique developed by Kaiser \& Squires
(1993). The projected maps of the total cluster mass are presented with a
relative normalisation and are thus {\it independent of any uncertainties in
the source $N(z)$}.  The unique feature of our study is the comparison
of these maps with the distribution of {\it both} cluster members and
the X-ray gas. We can therefore compare the large scale distribution of
dark matter with that for the baryonic component in two clusters. Our
study is thus a logical extension of Tyson {\it et al.}'s original
method. By creating a deeper sample of background galaxies and applying
a new analytic technique, we can achieve a sufficiently high galaxy
surface density in good seeing to allow us to resolve structure in the
lensing clusters.

A plan of the paper follows. In Section 3 we begin by briefly recapping
the target selection, optical data acquisition and reduction methods
before discussing the reduction and analysis of the ROSAT X-ray images
of the two selected clusters. In Section 4 we discuss our
implementation of the Kaiser \& Squires procedure. The mass maps are
presented in Section 5 and compared with the distributions of
galaxies and hot X-ray gas in the clusters.  Section 6 gives our
discussion of the comparison between the mass distribution and that of
the baryonic tracers. Our conclusions are summarised in Section 7.

\section{OBSERVATIONAL DATA AND METHODS}

The philosophy behind our observational programme is thoroughly discussed in
Paper I where the criteria by which the clusters were selected are given.
All of the optical CCD observations and their reduction procedures are
reviewed in considerable detail. Here we briefly summarise the optical
characteristics and respective datasets for the two clusters for which mass
distributions are presented before discussing the X-ray images and their
treatment. The latter data was not presented in Paper I.

The 3 clusters chosen for our survey span the redshift range
$0.26<z_{cl}<0.89$ and X-ray luminosities are available for each. 1455+22
($z_{cl}=0.26$) was originally identified by the EINSTEIN X-ray
satellite (Henry \et 1992) whereas 0016+16 ($z_{cl}=0.55$) was first
found on deep optical photographic plates (Koo 1981). Both clusters are
amongst the most X-ray luminous examples known and can, effectively, be
regarded as X-ray selected. The highest redshift cluster, 1603+43
($z_{cl}=0.89$) was optically discovered and is a much weaker X-ray emitter.
No significant lensing signal was detected with this cluster, which
provides important constraints on the redshift distribution of the
field galaxies (Paper I) but clearly eliminates it from further
consideration in any determination of cluster mass distributions.

The optical observations were made in July 1990 and May 1991 using the
TAURUSII f/4 focal reducer on the 4.2m William Herschel Telescope (WHT)
(see Table~2 of Paper I) with the largest format EEV CCD then
available. This arrangement provides  0.27 arcsec pixel$^{-1}$
sampling over a 5$\times$5 arcmin field (corresponding to 1.5 Mpc and
2.2 Mpc for the clusters concerned).  Multiple exposures of duration
$\ls$ 1000 sec were taken in $V$ and $I$ offsetting in-between each so
that the data frames can be combined to create a sky flatfield for the
entire night. The $I$ band data was used to determine image shapes for
the lensing analysis. For 1455+22 the final $I$ frame has an effective
seeing of 0.90 arcsec FWHM, for 0016+16 it is 0.95 arcsec FWHM.  The
final on-source integrations for the two clusters are: 1455+22 --
$t_{exp}(V)$=12.0 ksec and $t_{exp}(I)$=20.8 ksec; 0016+16 --
$t_{exp}(V)$=11.0 ksec and $t_{exp}(I)$=25.5 ksec.

The dataset was reduced to matched $V,I$ object catalogues using the
FOCAS image processing algorithm (Jarvis \& Tyson 1981) with some minor
modifications as described in detail in Paper I. The final catalogues
have 1 $\sigma$ isophotal limits of: 1455+22 -- $\mu_V = 28.9$ mag
arcsec$^{-2}$ and $\mu_I = 27.8$ mag arcsec$^{-2}$; 0016+16 -- $\mu_V =
28.8$ mag arcsec$^{-2}$ and $\mu_I = 28.2$ mag arcsec$^{-2}$.  These
limits are sufficiently faint to obtain reliable ellipticities and
colours to at least $I$=25.  At this limit, the signal to noise of a
single galaxy image is typically $\simeq$15-20$\sigma$ in the seeing
disk.  The 80\% completeness limits for the catalogues are $I$=25.3 and
$V$=26.5 for 1455+22 and for 0016+16, $I$=25.7 and $V$=26.4.  The high
source densities available ($\sim$40 arcmin$^{-2}$ at $I$=25) enables
us to resolve mass structures on scales comparable to those available
from our X-ray images.  Note that although a lensing signal could be
{\it detected} using a brighter magnitude limit with fewer field
galaxies per unit area, very deep images are required to derive mass
maps of the required spatial resolution {\it and} signal to noise.

We now describe the properties of the individual clusters, paying
particular attention to the X-ray images and their reduction.

\subsection{1455+22}

This cluster was discovered as a serendipitous source in the EINSTEIN
Medium Sensistivity Survey (EMSS, Henry \et 1992).  The cluster is the
most X-ray luminous cluster within a redshift of 0.5 in the EMSS: $1.6
\times 10^{45}$ ergs sec$^{-1}$ in the 0.3--3.5 keV band.
Unfortunately, redshifts are available for only 4 cluster members
(Mason \et 1981), including the dominant central galaxy ($z=0.258$).
The velocity dispersion determined from so few galaxies is of course
highly uncertain, $\sim700^{+2000}_{-100} $kms sec$^{-1}$, where the
uncertainties are 90\% confidence limits assuming a gaussian velocity
dispersion.  The central galaxy also exhibits the strongest optical
line emission of any of the EMSS clusters (Donahue \et 1992) and
therefore probably contains a cooling flow.  While the X-ray luminosity
is very high the optical richness derived from counts of
colour-selected cluster members (see below) is low, being about half
that expected from lower redshift clusters (Edge \& Stewart 1991) and a
third of the value for Coma.

As a part of a larger programme to obtain detailed X-ray images of a
sample of moderate redshift clusters ($z=$0.2--0.3) for X-ray/optical
gravitational lensing study, a ROSAT HRI observation was taken of
1455+22.  A total exposure of 8.3 ksec was obtained in two parts on
11$^{\rm th}$January 1992 and the 20$^{\rm th}$January 1993.  Although
this exposure is half that originally requested it provides sufficient
signal-to-noise to detect the cluster out to 1~Mpc and resolve the
central 50~kpc of the cluster. Figure~1(a) shows this image overlaid on
the composite V+I WHT image. The emission is highly peaked on the
central galaxy but shows no evidence for a significant point source
(less than 5\% of the total X-ray flux). Deprojection of the surface
brightness profile (Arnaud 1987, White 1992) indicates that the cluster
does indeed contain a cooling flow of 630$^{+257}_{-178}$ \Msunpyr.
The deprojection analysis uses an assumed form for the cluster
potential and combines this with the observed X-ray surface brightness
distribution to determine a temperature profile for the cluster.
Ideally this is then matched to the observed temperature profile.  In
the absence of such information this procedure unfortunately prevents
the determination of an exact mass profile.  However, taking an
isothermal temperature of 8$\pm3$ keV for the cluster (consistent with
the luminosity-temperature relation) and a core radius where the ratio
of gas mass to total mass is constant, the best fit King potential has
a velocity dispersion of 1000$\pm$200 kms sec$^{-1}$ and a core radius
of 150$^{+100}_{-50}$ kpc.  A more detailed description of the analysis
and how the results relate to other clusters will be presented
elsewhere (Edge \et in prep.).

In summary, the high inferred mass and its compact distribution
strongly suggest that this cluster is  a very good candidate for deep
lensing studies proposed in Section 1.

\subsection{0016+16}

This cluster is the second most luminous EMSS cluster (Henry \et 1992)
at $1.43 \times 10^{45}$ ergs sec$^{-1}$.  However, in contrast to
1455+22, it has been well studied both optically and in the
near-infrared (Koo 1981, Ellis \et 1985, Arag\'on-Salamanca \et 1993).
The cluster's redshift is $z = 0.545$ with a rest-frame velocity
dispersion of $\sigma = 1324$ kms sec$^{-1}$ from 30 members (Dressler
\& Gunn 1992, and priv.\ comm.).

Morphologically, the cluster is very different to 1455+22 containing no
dominant central galaxy. The peak in the galaxy surface density lies
slightly to the south-west of a linear structure defined by 3 bright members
(Figure~1(b)). The optical counts indicate a richness of $\simeq 2\times$Coma,
but possible contamination by foreground systems may reduce this estimate
(c.f.\ Ellis \et 1985). The absence of a prominent population of blue members
suggests the cluster is unusual advanced in evolutionary terms for
its redshift.  This conclusion is supported by the analysis of the
rest frame \UV\ colours of the red early-type cluster members.  These
exhibit a remarkably small scatter showing that the population is
extremely homogeneous and implying that it is very old even at $z=0.55$.

A deep EINSTEIN HRI image of 0016$+$16 (White \et 1981) determined an
X-ray core radius of 220 kpc which is comparable to the intrinsic
resolution of the image. Unfortunately the high internal background of
the EINSTEIN HRI prevented any study of more extended, low surface
brightness emission.  Recently a ROSAT PSPC image with an exposure of
43.2 ksec was obtained by Hughes and collaborators, this is now
publicly available from the ROSAT archive. The hard (0.4--2.4~keV)
image is overlain on the composite V+I frame in Figure~1(b).
The X-ray map reveals an elliptical or bimodal distribution centred on
the optical cluster center. A detailed comparison of the X-ray and
optical morphologies is given below.  A deprojection analysis of the
X-ray surface brightness profile with a King model potential gives a
velocity dispersion of 1300$\pm$200 kms sec$^{-1}$ and a core radius of
400$^{+200}_{-150}$ kpc.

0016+16 is a very rich and concentrated cluster and ideally suited for
lensing studies. In addition its morphology contrasts usefully with
that of 1455+22.

\subsection{Field and Cluster Galaxy Catalogues}

Paper I describes in more detail how photometric catalogues of galaxies
to $I\ls25$ can be used to define `field' and `cluster' samples on a
statistical basis. From the aperture \VI\ colours, well-defined
colour-magnitude relations are observed for the early-type members of
both clusters. The tightness of these relations ($\Delta(V\!-\!I) = 0.04$
mag for 1455+22 and $\Delta(V\!-\!I) = 0.06$ mag for the bright end of the
0016+16 sequence), enables us to selectively label galaxies in the
sequence as `cluster members' and those outside this narrow colour
relation are assumed to be `field galaxies'. The number-magnitude
counts for this field population agree closely with published data in
random fields at high Galactic latitude (Lilly \et 1991). The same
procedure can be used to show there is no statistically significant
excess from cluster members fainter than $I$=22.0 (for 1455+22) and
$I$=23.5 (for 0016+16).  These limits are equivalent to absolute
magnitudes of $M_V \sim -17$ and $M_V \sim -18.5$ respectively.

For 0016+16, an excess of {\it bright} galaxies is
apparent, most likely associated with foreground groups identified by
Ellis \et (1985). For this cluster, additional colour information is
available from an $R$ band Service exposure taken with a large format
EEV CCD at the 2.5m INT prime focus.  This frame has a total exposure
time of 6 ksec and is adequate to provide colours to $I$=24 accurate to
better than 0.2 mag. Additional colour criteria were thus used on the
(\VR)--(\RI) plane to isolate galaxies with colours similar to E/S0's
at $z=0.55$.

In the case of 1455+22, our cluster samples contains $\simeq$180
galaxies brighter than $I$=22 over the 5.4$\times$5.0 arcmin field. For
0016+16, we have 174 cluster galaxies to $I$=23.5 in a 3.3$\times$5.1
arcmin region.  The corresponding field samples contain 1583 and 831
galaxies respectively above their 80\% completeness limits.

We note that neither of the clusters presented here show giant arcs
($a/b\ge10$) down to surface brightness limits of $\sim$0.05\% of the
night sky.  The absence of giant arcs in these two clusters is not a
particular concern as their creation is very sensitive to the detailed
structure of the mass distribution on small scales in the cluster
centres (e.g.\ Smail \et 1991).  As we show below, both our clusters
{\it are} massive lenses and therefore the absence of giant arcs even
to very deep limits has little or no bearing on this.   This conclusion
compromises the use of shallow surveys for giant arcs to study the
properties of the lensing clusters, although such surveys are still
useful to study the high redshift galaxies seen as arcs.

The lensing techniques described below rely upon our ability to
accurately estimate the ellipticities of faint objects.  This issue is
discussed at length in Paper I where we argue for the introduction of a
new weighting scheme for ellipticity measurements compared to that
originally implemented in the FOCAS package. Our approach, originally
proposed by Bernstein (priv.\ comm.), is to use a {\it radial}
weighting function within {\it circular} apertures when calculating
second moments instead of using the detection isophote to define pixel
membership at each surface brightness. In this scheme, the optimal
weighting function  has the same profile as that for the image itself.
Bernstein has shown that using profile shapes individually tailored to
each object does not significantly improve the weighting scheme
considering the large computational burden introduced (Bernstein
priv.\ comm.).  The weighting function adopted therefore is a circular
gaussian whose variable width is determined from the intensity-weighted
radius of the image after seeing convolution. We refer to these moments
as `optimally-weighted'.

Using tests discussed in Paper I, we demonstrated a considerable
improvement in the robustness of the optimally-weighted ellipticity
measurements for very faint objects compared to the traditional
measurement. For a typical $I\simeq25$ galaxy we obtain roughly a
four-fold reduction in the scatter:  ($<\!\!  \Delta\epsilon_{opt}
\!\!> = 0.04$ versus $<\!\!  \Delta\epsilon \!\!> = 0.16$  from
analysis of independent frames of the same field). However, tests of
simulated frames populated with objects of known ellipticity indicate
this improvement in reliability is at the expense of a small systematic
bias (rounder by $\simeq0.1$) in the returned ellipticity (where the
FOCAS moments provide an unbiased measure). As we are primarily
interested in the {\it relative} strength of the lensing signal across
our field, the disadvantage of a small systematic bias is outweighed by
the improved signal/noise of the optimally-weighted moments.  In
particular we can adopt a lower effective ellipticity cut-off than when
using FOCAS moments as the orientations of rounder
objects can be successfully measured due to the lower scatter.  Owing
to the intrinsic distribution of image shapes this provides a much
larger sample of objects for analysis.

\section{DETERMINING THE DISTRIBUTION OF DARK MASS}

Various statistical tools have been developed to analyse the weak
lensing of faint galaxies by rich clusters. The methods fall into two
main classes:  parametric likelihood tests which assume some form for
the distribution of mass in the lens and then attempt to determine the
most likely values of the model parameters (Kochanek 1990,
Miralda-Escud\'e 1991a, 1991b, Paper I), and non-parametric tests which
directly derive the mass distribution from the variation of the lensing
signal across the cluster image (Kaiser \& Squires 1993).

The parametric methods are particularly well-suited for constraining
the faint galaxy redshift distribution provided there are some
constraints on the depth and scale of the cluster potential well, e.g.\
from dynamical or X-ray data. In Paper I we used likelihood versions of
these methods to obtain the most probable redshift distribution for the
$I$=25 field population from the lensing signal observed in the 3
clusters at $z_{cl}$=0.26, 0.55 and 0.89.

Alternatively, the non-parametric technique developed by Kaiser \&
Squires (1993, see also Fahlman \et 1994) is better suited for
investigating the relative distribution of mass in the lensing cluster.
Not only can the method provide a genuine `map' of the projected mass
distribution in moderate and high redshift rich clusters with a
resolution appropriate for comparison with the baryonic tracers, but
importantly, the results are totally independent of the {\it in situ}
estimators allowing us to test the basic assumptions those methods
adopt.

The mathematical derivation of the projected mass density estimator
in the Kaiser \& Squires (KS) method is not repeated here; the interested
reader is referred to their original article. The basic principle is
that the distortion signal produced by a foreground point mass is of a
fixed pattern. This pattern can be compared to the observed alignment of
faint galaxies (positions $\vec r_g$) around a selected point ($\vec
r$) in the cluster image plane. The degree of similarity between the
two patterns is a direct estimate of the surface density of lensing
mass at that point, $\Sigma({\vec r})$. The statistic is evaluated
repeatedly over a grid of centres (40$\times$40) across the cluster
yielding a `map' of the projected  mass.   We define $e_1$ and $e_2$
such that $e_1$ measures the stretching of a galaxy image along the X and
Y axes while $e_2$ measures the stretching in the direction Y=X
in terms of the intensity weighted second moments of the image
shape, $I$.
$$ e_1 = \left( \frac{I_{xx}-I_{yy}}{I_{xx}+I_{yy}} \right) \qquad
   e_2 = \left( \frac{2I_{xy}}{I_{xx}+I_{yy}} \right)$$

Kaiser \& Squires show that the surface mass density, $\Sigma({\vec r})$,
is related to the local induced distortion by the
sum over the components $e_1$ and $e_2$ for all galaxies around that position
weighted by the function, $W({\vec r_g} - {\vec r})$.
$$ \Sigma({\vec r}) = \frac{1}{\overline n}
\sum_{\rm galaxies} W({\vec r_g} - {\vec r}) \;
\chi_i({\vec r_g} - {\vec r}) \; e_i ({\vec r_g}) $$

$$ \chi_1({\vec r}) = \frac{(x^2 - y^2)}{r^2}
\qquad \chi_2({\vec r}) = 2 \frac{ (xy)}{r^2} $$

$W({\vec r_g} - {\vec r})$ is the  pattern of the induced distortion
from a point mass as a function of radius.
The average surface density of background sources in the field is given by
$\overline n$.

The reconstruction is unstable to the noise caused by the intrinsic
orientations and ellipticities of the background sources. To overcome
this   the derived mass distribution is filtered.  The filter function,
$T(r)$, can be directly combined into the pattern function, $W(r)$, for the
point mass template.
$$ W(r) = \frac{1}{2} \int^\infty_0 \kappa  \; T(\kappa)  \;
(2 J_1(\kappa r)/\kappa r - J_0(\kappa r))  \; d\kappa $$

In the following analysis we use a gaussian for $T(r)$, with a
separate $\sigma$  adopted for each cluster.
An estimator for the uncertainty in the mass reconstruction is also given by
Kaiser \& Squires:
$$ \langle \Sigma({\vec r}) \rangle = \frac{1}{8 \pi^2 \overline n} \;
\langle e^2 \rangle  \int^\infty_0 \;  T^2(\kappa) \;  d^2\!\kappa $$

where $\langle e^2 \rangle$  is the intrinsic dispersion in the
ellipticities of the background galaxy population.  For a given filter
function the errors depend strongly on the surface density of sources
(hence the importance of obtaining very deep images) and weakly on
their redshift distribution $N(z)$ assuming the cluster is foreground
to the bulk of the population.  The systematic effects of our weighting
scheme used in the ellipticity measurement are compensated for by the
factor $\langle e^2 \rangle$.  The form of the error estimator given
above is only valid for a dataset of infinite extent. In reality the
error depends upon the number of objects contributing, through the
filter function, to the reconstruction at each grid point.  Close to
the frame border the average weighted sum over the contributing images
will be less than in the frame centre by a factor dependent upon the
form of the filter function.  The resulting variable signal to noise
across the reconstruction makes simple mass maps hard to interpret. In
the following discussion, the mass maps produced from the standard KS
prescription are further divided by a map of the noise estimated using
a local source surface density.  The resulting maps have constant noise
properties and can be more easily visualised. Apart from the
morphological comparison our other analysis uses the standard KS mass
map.  The local noise estimator is:
$$ \langle \Sigma({\vec r}) \rangle
= \frac{1}{8 \pi^2 \overline n} \;
\langle e^2 \rangle \; \sum_{\rm galaxies} \;  T^2(\theta)  $$
Where ${\vec \theta} = {\vec r_g} - {\vec r}$.

We can make a simple test of the KS noise estimator and the systematics
in our data by analysing the deep images of our third cluster 1603+43,
$z=0.89$, discussed in Paper I but not included in this work because
significant lensing is not detected even with low order statistics such
as the orientation histogram (Figure 9(c) of Paper I). We therefore
adopt the cluster as a `blank field'.  Using the global error estimator
the highest significance peak in the reconstruction is 3.4 $\sigma$ on
the edge of the frame.  Using the local error estimator this spuriously
high significance is reduced to 2.5 $\sigma$.  This value is broadly
consistent with expectations from random noise given the number of grid
centres used.  The quoted errors were estimated using the no-evolution
$N(z)$ for the sample magnitude limits ($I \in [23,25.5]$) as discussed
in Paper I.

When estimating the total mass in our clusters we have to correct the
values taken from our mass maps as the surface density estimator as
given assumes that the surface density is zero at the frame boundaries
(Fahlman \et 1994). At the cluster radii available in our field of view
this is not true and we should therefore correct the central mass
estimates for the mean mass surface density at the frame border.  This is
unfortunately ill-defined, although using a King profile for the
cluster mass we would anticipate that the correction ought to be small,
only $\sim$20\% of the central mass.   Given the additional
uncertainties due to the background redshift distribution we prefer to
include this systematic error in the overall error budget for our mass
estimates.

\section{RESULTS}

In this section we present the mass maps for 1455+22 and 0016+16 using
the Kaiser \& Squires technique modified as discussed in $\S$4.
Ellipticities and orientations of the field samples used the optimal
weighting scheme discussed in $\S$3 and Paper I. The sample magnitude
limits are:  $I \in [23,25.3]$ for 1455+22 and $I \in [23,25.5]$ for
0016+16.  These limits provide the highest  surface density of sources
with reliable shape measurements whilst minimising the foreground
contamination (for our adopted redshift distribution). All objects
whose isophotes touch the frame boundaries have been removed from the
cluster catalogues.

\subsection{1455+22}

In Figure~2 the mass map derived from the Kaiser \& Squires analysis
of the field catalogue is compared with the smoothed number density
distribution of colour-selected cluster members defined according to
the prescription discussed in $\S$3.  Adopting a straight number
weighted scheme is equivalent to the assumption that the galaxy's
luminosity does not appreciably effect its dynamical properties.  For
both maps the smoothing scale is 135 kpc (0.45 arcmin) and is shown by
a scale bar. The mass contours intervals are based on errors calculated
using the local weighting scheme discussed in $\S$4 thereby reducing
spurious features on the frame boundary. These errors also depend on
the adopted redshift distribution, $N(z)$, for the faint galaxies.
Following Paper I, we adopted the `no evolution' $N(z)$ for both
cluster analyses, but our results are not sensitive to this
assumption.  Indeed, the {\it relative} mass maps are independent of
$N(z)$.

The morphological similarities between the distributions of cluster
members and lensing mass are striking.  If we fit elliptical contours
to the two distributions  out to a scale of 360 kpc (1.25 arcmin), the
position angles ($\theta$) agree within the errors:   $\theta_{gal} =
145 \pm 2$ degrees compared to  $\theta_{mass} = 146 \pm 2$ degrees.
This is illustrated in Figure~3 for all four distributions available
for 1455+22.  More interestingly the ellipticities ($\epsilon$) of the
two distributions are also in surprisingly close agreement
$\epsilon_{gal} = 0.52 \pm 0.03$ and $\epsilon_{mass} = 0.47 \pm
0.03$.  This is as expected if the galaxies act as virialised mass-less
tracers of the cluster potential.

To determine if the elliptical mass distribution is
produced by unresolved substructure reconstructions were performed
using smaller smoothing scales.  While noisier the mass peak does not
show any internal structure on scales greater than 60 kpc.  In support
of this on yet smaller scales ($\ls$40 kpc), the orientation of the cD
($\theta  = 148\pm2$ degrees) is also a close match to the mass
distribution as would be expected from dynamical arguments that account
for its growth.  The ellipticity of the cD halo on these scales is
$\epsilon_{cD} = 0.23 \pm 0.01$, considerably rounder than the mass or
cluster galaxy distributions.  These parameters are summarised in
Table~1.

The agreement between the positions of the peaks of the
distributions is less good, although the discrepancies while formally
significant are all less than the respective smoothing scales.  The
offset between the peaks in the galaxy number density and mass
distributions is 120$\pm$20 kpc, with the cD offset 90$\pm$15 kpc from
the peak of the galaxy number density.  Both the mass and galaxy number
density distributions peak to the north-west of the cD. These offsets
are a concern but we note that there is growing evidence that
cD's do not always lie at the dynamical centres of their cluster
(e.g.\ Bird 1994).
The positional offset determined here is negligible compared to
that inferred in other systems from the observed velocity offsets.
Also, for a cluster with asymmetry
along the line-of-sight it is not necessary for the position of the
projected peak surface density to exactly match the projected position
of the peak in the local density distribution.

Figure~2 also shows an equivalent comparison between the mass map and the
X-ray surface brightness distribution smoothed to the same resolution,
the X-ray surface brightness is plotted with a logarithmic intensity
scale.  The X-ray surface brightness peaks within 40 kpc of the cD.
Again the contours are similarly oriented with $\theta_{X} = 136 \pm 8$
degrees and $\epsilon_{X} = 0.16 \pm 0.06$ out to 360 kpc.  On the
assumption that the X-ray surface brightness reflects the shape of the
the potential then, for a logarithmic potential at radii $r \gg r_c$,
we can convert the ellipticity of the X-ray contours into that of the
surface density (Binney \& Tremaine 1987, Eq.\ 2-55).  This gives
$\epsilon_{mass}^X \simeq 0.46\pm0.18$, in very good agreement with the
value determined from fitting to the mass surface density.

Whilst there is obviously excellent agreement between the projected
morphologies of the three distributions, in terms of shapes and
orientations, it is critical to know whether the lensing matter has the
same characteristic scale length as that of the visible baryonic
component.  Figure~4 shows the radial profiles derived from the mass
and red cluster galaxy distributions.  Both profiles have been centred on
their respective maximum, corrected for the ellipticity of their
distributions and normalised to the inner-most bin. The error-bars show
the spread in values at a given radius.  For clarity the galaxy profile
has been offset slightly.

Within 400 kpc (1.3 arcmin) both profiles obey a similar functional
form, with the distribution of galaxies being marginally more
extended.   To quantify this we fit a modified Hubble profile
($\Sigma(r) \propto 1/(1+(r/r_c)^2)$) to the projected distributions.
The adopted functional form is an adequate description of the shapes of
our distribution profiles given the limited range.  For comparison the
parametric likelihood analysis undertaken for 1455+22 in Paper I gave a
best fitting core radius of $r_c^{mass}=210\pm100$ kpc.  Here we obtain
maximum likelihood values of: ($r_c$):  $r_c^{mass} = 200^{+70}_{-40}$
kpc and $r_c^{gal} = 210^{+80}_{-40}$ kpc respectively. These values
are azimuthally averaged and retain the effects of the smoothing used
to generate the distributions.  These are very similar to the core
radii estimates obtained from X-ray imaging: $\sim$200 kpc (c.f.\ Jones
\& Forman 1992).  The X-ray core radius from analysis of our HRI image
is $r_c^{X} =150^{+100}_{-50}$ kpc.  We can obtain the true core radius
by correcting the profiles for the effects of the smoothing function
and the ellipticity of the observed distributions.  Comparison of the
observed distributions with model profiles convolved with the same
smoothing function gives maximum likelihood fits of $r_c^{mass} =
100^{+80}_{-50}$ kpc and $r_c^{gal} = 180^{+70}_{-50}$ kpc corrected to
the semi-major axis.  Within the errors from our model fitting we
cannot formally discard the hypothesis that the two distributions have
the same core radius. However, the more extended scale of the galaxy
distribution is clearly apparent in Figure~4 at radii $r \gs 200$ kpc.

While a useful exercise, comparing parameters derived from functional
fits to the 1-D profiles is a relatively insensitive tool for
determining if the distributions have different scale-lengths.  A more
informative approach to understanding the relative distribution of mass
and light is to take the 2-D maps and calculate the effective
`mass/light' ratio as a function of position.  Figure~5 shows the
median ratio of the mass surface density ($\Sigma_{mass}(r)$) to the
galaxy surface density  ($\Sigma_{gal}(r)$) as a function of radius in
the cluster taken from this comparison: $M/N_{gal} = \Sigma_{mass}(r) /
\Sigma_{gal}(r)$. The error bars denote the 1 sigma scatter at a given
radius and the points are normalised to the central bin.  This ratio
shows a constant decline out to 500 kpc, amounting to a factor of 3
drop.  Figure~5 thus confirms the earlier suggestion, based
on the profile fitting,
that the mass distribution is more centrally concentrated than the
galaxies, the observed trend is significant at the 3.3 $\sigma$ level.
Adopting a luminosity-weighting scheme for the galaxy distributions
rather than our number-weighted one gives a marginally flatter, but
statistically indistinguishable, slope for $M/N_{gal}$.  This
luminosity-weighted scheme has a larger scatter but indicates a drop of
a factor of 2 in the mass/light ratio out to 500 kpc, although the
slope is also consistent with zero within the large errors.

Returning to Figure~2 we note a significant secondary maximum
($\simeq 10 \sigma$) in the mass map $\simeq$ 500 kpc due east of the
cluster centre. As this is well clear of the frame boundary, it is
important to determine if this peak correlates with any other physical
feature. Interestingly the peak does not lie close to any feature in
either the distributions of X-ray emission, cluster members or field
galaxies.  The bow-wave like structure seen in the mass map persists
when we use a smoothing scale of 60 kpc in the reconstruction, showing
that the structure does not consist of individual mass peaks on scales
larger than about 60 kpc.

Turning to the X-ray analysis, we have shown that the potential adopted
is reasonably representative of that delineated by the lensing mass.
We can then combine the total cluster mass within a radius of 450 kpc,
derived from the X-ray analysis on an {\it absolute} scale with the
lensing signal to estimate the mean redshift of the $I\leq25$
background sources. This can be viewed either as a check of the
conclusions of Paper I, or, adopting $N(z)$ from Paper I, it provides a
self-consistent check on the X-ray mass estimates.

Figure~6 shows the variation in the cluster mass derived from the
lensing analysis parametrised as a function of the median source
redshift.  This is a lower limit to the mass as the KS technique
assumes the mass surface density goes to zero at the boundary of our
5$\times$5 arcmin field (c.f.\ Fahlman \et 1994).  Also shown is the
the projected mass determined from the X-ray analysis.  Due to
extrapolation to large radii necessary to determine the {\it projected}
mass from the X-ray analysis it would be better to deproject the
lensing mass distribution to give the mass of the central regions of
the cluster and directly compare this with the X-ray determination.
Unfortunately this deprojection requires the cluster mass profile at
large radii which currently is poorly-constrained observationally.  The
flatness of the curve combined with the probable errors in the X-ray
mass estimate make this a relatively insensitive tool with which to
derive an upper limit to the background source redshift. A spatially
resolved temperature profile for 1455+22 would significantly reduce the
systematic errors in the X-ray mass determination.  Nevertheless,
Figure~6 does provide a rigorous lower limit of $M \gs 2.2 \times
10^{14} M_\odot$ for the projected cluster mass in the central 0.9 Mpc
or so -- comparable to $M_X \ls 2 \times 10^{14} M_\odot$ in the
similar region of Coma from Hughes (1989) most concentrated X-ray
models.  Moreover, if we adopt the preferred redshift distribution from
Paper I we obtain good agreement between the X-ray and lensing derived
total masses in the cluster centre.  The resulting mass in the cluster
centre is $M \sim 3.1 \times 10^{14} M_\odot$, roughly 50\% higher than
the maximum equivalent value for Coma.  We note that this mass estimate
is independent of the dynamical state of the cluster.

Studying Figure~6 it is readily apparent that
a difference between the X-ray and lensing determined masses of the
magnitude claimed by Babul \& Miralda-Escud\'e (1994) if
typical of the whole cluster would be
difficult to achieve unless the field redshift distribution was peaked close
behind 1455+22.  As we will see below this can be ruled out by a
similar analysis of 0016+16 which also shows good agreement between the
central cluster masses derived from the lensing and X-ray data.

Having determined that there is relatively good agreement between the
mass estimates from the lensing and X-ray analyses we convert these
into rough estimates of the mass to light ratio for the cluster.  The
integrated luminosity of all the red cluster galaxies within 450 kpc of
the cluster centre  is $L_V = 6.8 \times 10^{11} L_\odot$.  This is a
lower limit to the total cluster luminosity owing to possible
contributions from foreground structures and cluster members with
colours bluer than the E/S0 sequence.  Integrating the light of all the
galaxies in the field seen within 450 kpc of the cluster centre
assuming they are cluster members gives $L_V \sim 1.0 \times 10^{12}
L_\odot$.   Using the lower bound to the cluster mass determined above
we can thus derive a minimum mass to light ratio for the cluster of
$M/L_V \gs 220$ in solar units.  This is similar to values derived from
virial analysis of rich clusters, $M/L_V \gs 250$.  However, adopting
the total central mass indicated by {\it both} the X-ray and lensing
methods and the red cluster galaxy luminosity requires $M/L_V \sim
460$.  For comparison the mass to light ratio required for closure
density is $M/L_V \sim 700$.

\subsection{0016+16}

This cluster is morphologically quite different to 1455+22 having no
central cD and a considerably higher optical  richness although with a
similar X-ray luminosity.  Nevertheless, most of the conclusions
derived in $\S$5.1 apply equally well to 0016+16.  The analysis of this
cluster is complicated primarily by the small field available,
the source density beyond $z=0.55$ is still sufficiently high at $I\sim25$ to
allow us to robustly map the cluster morphology.

Figure~7 compares the lensing mass map derived from the Kaiser \&
Squires technique with the distribution of red cluster members.  The
effective spatial resolution in the mass reconstruction is 200 kpc
(0.45 arcmin). Although the salient features reproduce between the two
maps, there are many more apparently spurious features in the lensing
maps away from the cluster centre than observed in 1455+22.
Nevertheless, the central region free of these features still comprises
an area equivalent to that probed in our intermediate redshift cluster,
1455+22.   The most striking feature common to both maps is a
elliptical-like peak with bimodal substructure straddling the
optically-defined centre. The mass map indicates the two clumps
have a projected separation of 600 kpc with the more concentrated
sub-clump in the south-west. The galaxy distribution reveals
at least 3 sub-clumps orientated similarly to the mass distribution.
A fourth sub-clump to the south-east is not detected in the mass map.
The orientations of the mass and galaxy distributions between 300--600
kpc agree closely: $\theta_{mass} = 131 \pm 6$ degrees and
$\theta_{gal} = 124 \pm 8$ degrees. As in 1455+22 the orientation of
the central galaxies (in this case a linear chain rather than the cD
envelope) follows that of the mass on larger scales ($\theta = 125 \pm
10$ degrees). The best fit ellipticities, shown in Figure~8, are
$\epsilon_{mass} = 0.59 \pm 0.01$ and $\epsilon_{gal} = 0.21 \pm 0.02$.

In Figure~7 we also compare the mass map with the ROSAT PSPC image of
similar spatial resolution, the latter displayed with a logarithmic
scaling.  Again the distributions are well aligned with $\theta_{X} =
127 \pm 4$ degrees over the range 300--600 kpc. The ellipticity on
these scales is $\epsilon_{X} = 0.21 \pm 0.02$. As for 1455+22 we
convert this to that appropriate for the surface density yielding
$\epsilon_{mass}^X \simeq 0.61 \pm 0.06$ for $r \gg r_c$ and
$\epsilon_{mass}^X \simeq 0.28 \pm 0.03$ for $r \ll r_c$. Adopting the
core radii determined below we are closer to the $r \gg r_c$ regime on
the scales where the ellipticity is measured.

Summarising, there is reasonable agreement between the ellipticity of
the mass distribution and that for the X-ray gas. Although the complex
structure in the galaxy distribution precludes a detailed comparison,
both the galaxy and mass distributions have similar bimodal forms. To
determine if the two peaks of the mass distribution are dynamically
distinct, we divided Dressler \& Gunn's (1992) spectroscopic sample
along a line perpendicular to the axis of the two sub-clumps. The
rest-frame velocity difference is $\sim 400$ kms sec$^{-1}$ and not
statistically significant.  In view of the apparently different surface
densities for the two mass peaks we would predict a temperature
difference between the X-ray gas in the two structures.  Therefore a
spatially resolved X-ray temperature map of the cluster from ASCA is of
great interest.

Figure~9 shows the projected profiles for the mass and galaxy
distributions centred on their respective maxima and normalised to
their innermost bin.  The galaxy distribution reveals an intrinsic
deconvolved core of $\sim$ 330 kpc (semi-major axis) after correction
for the ellipticity of the galaxy distribution.  This is larger than
the value obtained from the mass distribution of $\simeq$ 210 kpc
although both values are systematically uncertain due to both the
complex structures in the two distributions and the problems of
background subtraction associated with the small field.  Nevertheless,
the mass core radius from our lensing reconstruction compares well with
that determined from the parametric analysis in Paper I,  $r_c^{mass} =
210\pm250$ kpc.  Equally the lensing core radius is smaller than that
derived from the PSPC image of 400$^{+200}_{-150}$ kpc.

Figure~10 shows the estimate of the total mass within a radius of 600 kpc of
the centre of 0016+16 as a function of the median redshift of the
background population.  The derived mass surface density is a lower
limit as it assumes that the project surface density in the cluster is
zero at our frame boundaries.  From the lensing analysis we determine a
robust lower limit to the projected mass of $M \gs  3.4 \times 10^{14}
M_\odot$.  This can be compared to the X-ray determined value for the
central 600 kpc of $M_X \sim  7.3 \times 10^{14} M_\odot$.  Adopting
the preferred median redshift for the background population, for the
magnitude range used in the lensing analysis, predicts a mass of $M
\sim  9.9 \times 10^{14} M_\odot$ close to the X-ray determined value.
As in 1455+22 we have good agreement between not only the morphology of
the mass determined from the lensing and X-ray analysis but also the
total masses inferred for the cluster.  For completeness we also
calculate the apparent mass to light ratio as for 1455+22 inside a
radius of 600 kpc.  Summing the luminosities of the red cluster members
in this aperture gives $L_V \sim 2.3 \times 10^{12} L_\odot$, while
summing all the galaxies in the field within this projected radius
yields $L_V \sim 4.9 \times 10^{12} L_\odot$  retaining the central
galaxies as the brightest cluster members.  Combining the lower limit
on the cluster mass from the lensing with the upper limit on the
luminosity we then obtain a mass to light ratio of $M/L_V \gs 70$ in
solar units, such a low value is not surprising given the large amount
of foreground contamination known to exist in this field.  Adopting the
luminosity of the red cluster members and the lensing derived mass we
calculate $M/L_V \sim 430$ in the central 1.2 Mpc of 0016+16.  The
uncertainties on this value are of course large.

\section{DISCUSSION}

When comparing the results from our two clusters we are struck by the
many common features observed, despite their different
morphologies and redshifts.

Firstly,  in each cluster we have four independent
estimates of the orientation of the major axis of the cluster projected
upon the plane of the sky (c.f.\ Table~1).  These four estimates span a
range in scales between the central galaxies ($\sim 20$kpc) out to
$\sim0.5 $Mpc (the lensing mass, the X-ray gas and the cluster galaxy
distribution).  In both clusters {\it all} four estimates are in good
agreement.  This implies that at least to first order the systems are
relaxed in their central regions.  This is the first time that
such a comparison has been made and it is encouraging to find such
good agreement.

Secondly, we find from the lensing analysis that both clusters have
moderately elliptical $\epsilon \sim 0.5$--0.6 mass distributions.
These values are close to the average ellipticity ($\epsilon \sim 0.5$)
predicted for clusters from the effects of tidal distortion on
proto-clusters by Binney \& Silk (1979).  Moreover, in both clusters
the ellipticities of the X-ray surface brightness distributions when
converted to the surface density agree with those derived from the
lensing signal.   This agreement does not extend to the galaxy distributions,
where in 1455+22 the galaxies act as tracers of the underlying mass
distribution, while in 0016+16 their general distribution is possibly
rounder and more clumpy. Thus it appears that in the more relaxed
system the galaxies are good tracers of the mass.  A similar conclusion
was reached for Abell 370 by Kneib \et (1993). Using detailed modelling of
several gravitationally lensed features they determine a mass
distribution in the cluster which closely matches the light.  We
believe our result is more robust as it is derived from the full two
dimensional mass and galaxy distributions rather than parametric fits
to these distributions.

{}From profile fitting in both clusters we find that the mass is
marginally more centrally concentrated than the two baryonic tracers,
although because of the large errors in profile fitting, this effect is
not formally significant. A direct comparison of the mass and galaxy
surface density distributions in 1455+22 gives a more significant
gradient in the ratio of mass to galaxy surface density. Specifically,
in Figure~5 we detect a factor of 3 drop in this ratio within $r<$500
kpc.  A similarly steep drop is claimed for the mass to light ratio in
the centre of the Perseus cluster (Eyles \et 1991).

The apparent concentration of mass compared to the baryonic tracers in
1455+22 is contrary to some theoretical predictions for hierarchical
growth of clusters (e.g.\ West \& Richstone 1988).  These show the
galaxies more centrally concentrated than the mass due to dynamical
friction.  One way to concentrate the mass relative to the galaxies
is for dissipation to occur preferentially in the mass component, this
appears unlikely if the mass is principally weakly interacting dark
matter, which is by definition incapable of dissipation.  Other ways
include removing galaxies from the central regions of the cluster (for
instance by merging them into a central galaxy) leading to an apparent
flattening of the galaxy surface density at smaller radii and
luminosity segregation, with fewer, but brighter, galaxies nearer the
centres of rich clusters.  Using luminosity-weighting for 1455+22 we do
indeed obtain a shallower slope for the variation of mass to light.
Unfortunately, due to the errors the slope is consistent with both zero
and that derived previously with number-weighting.  A further
possibility is to invoke incomplete virialisation to explain the
relative distributions of galaxies, gas and mass.  However, the close
morphological similarities between the distributions on large scales
would require us to have caught both clusters between the point where
they have relaxed sufficiently to allow the distributions to have
aligned but not quite enough for them to have reached equipartition.

With adequate data for only two clusters, it is clear that above option
cannot be ruled out.  A larger sample is needed to fully explore the
dynamical states of the centres of rich moderate redshift clusters.
Enlarging the angular coverage available for the lensing analysis will
also allow us to determine the asymptotic form of the radial mass
profile and compare this to that shown by the baryonic components of
the cluster.  This will provide a simple and direct test of the
existence of dark halos around clusters (Smail, Edge \& Ellis 1994, in
prep).

The  estimates of the total projected mass in the central 1-1.5 Mpc of
our two clusters from the X-ray analyses and the  lower limits
available from our lensing technique are in reasonable agreement.  If
we adopt the redshift distribution derived independently of the X-ray
analysis in Paper I this agreement becomes good: within 16\% in 1455+22
and 26\% for 0016+16.  Without detailed temperature information about
the X-ray gas in our clusters it is impossible to make a more detailed
comparison of the mass profile inferred from the X-ray imaging and that
provided by the lensing analysis. Nevertheless, adopting reasonable values
for the global cluster temperatures our result is inconsistent with an
extrapolation to larger scales of Babul \& Miralda-Escud\'e (1994)
result from their analysis of arcs in cluster cores.

Probably the easiest way to reconcile these two results is to recognise
that optical identification of clusters preferentially selects systems
with high projected central galaxy densities.  With a population of
reasonably high ellipticity clusters, as seems to be indicated both
from this work and optical studies (Plionis, Barrow \& Frenk 1991),
this bias will result in a large fraction of the more distant systems
having their major axis parallel to the line of sight.  Such a bias
will cause the projected mass surface density (i.e.\ along the major
axis) to be higher than the inferred mass from the X-ray analysis which
adopts a spherical symmetry.  Such an effect was discussed by Babul \&
Miralda-Escud\'e but discarded because they did not consider the
possible selection bias present in the original cluster
identification.  Two of Babul \& Miralda-Escud\'e's clusters are very
concentrated optically and may fall into the aligned class, the third
cluster, their most X-ray luminous, is not optically concentrated and
as they show has a lensing derived mass which is consistent with the
X-ray determination in accord with our findings.  Owing to their
geometry and the selection criteria adopted neither of our clusters
should be affected by this bias.  Hence, it appears that a more
detailed study of the biases inherent in studying optically-selected
clusters with giant arcs may hold the explanation to our disagreement
with the apparent results for cluster centres derived by Babul \&
Miralda-Escud\'e.

Combining our estimated masses with the luminosities of the red cluster
members gives mass to light ratios for the central regions of the
clusters of $M/L_V \sim 460$ for 1455+22 and $M/L_V \sim 430$ for
0016+16.  Correcting these for the surface densities at the frame
border assuming a King profile gives $M/L_V \sim 550$. If these regions
are representative of the Universe as a whole we obtain $\Omega \sim
0.8$.  We reiterate that these results are independent of the dynamical
state of the clusters.

Finally, studying both clusters it is apparent that they contain
significant substructure in their mass distributions.  We detect an
apparently significant sub-clump in 1455+22, while 0016+16 is strongly
bimodal.  This observation, if supported by a larger sample, has
interesting consequences for the growth of clusters and the evolution
of the galaxies within them.  Comparing the lensing maps with the X-ray
surface brightness (which follows the potential) highlights the
sensitivity of the lensing technique for studying the occurrence of
substructure in clusters.

\section{CONCLUSIONS}

Deep statistical lensing studies such as those presented here
are both difficult and time consuming. However, they present
one of the cleanest methods of studying the distribution of mass
in rich clusters.  The mass mapping technique of Kaiser \& Squires (1993)
is a very powerful and model-independent method for probing the
morphology of the cluster mass.

\noindent{$\bullet$} We have reconstructed projected mass distributions
for two luminous X-ray selected clusters, one at intermediate (1455+22;
$z=0.26$) and one at moderate redshift (0016+16; $z=0.55$). These are
accurate on a relative scale independent of any assumed redshift
distribution for the background field galaxies. Despite the different
cluster morphologies, both mass maps show a remarkable similarity to the
X-ray and galaxy surface density maps of the clusters on large scales.
This agreement extends to include the orientations and ellipticities of
the distributions.

\noindent{$\bullet$} A comparison of the relative distribution of the
mass and the galaxies shows that whilst galaxies are good tracers of
the mass {\it morphology} they appear to be less concentrated. However,
the X-ray gas and mass are more closely distributed. In our intermediate
redshift cluster we find a continuous decrease of $\times$ 2-3 in the
ratio of mass to galaxy surface density out to the limits of our data,
$r \sim 500$kpc. We obtain corrected maximum likelihood core radii for
the 3 distributions of $r_c^{mass} = 100^{+80}_{-50}$ kpc, $r_c^{gal} =
180^{+70}_{-50}$ kpc, $r_c^{X} =150^{+100}_{-50}$ kpc.

\noindent{$\bullet$} The moderate redshift cluster (0016+16) shows a
large amount of structure in the mass, X-ray gas and galaxy
distributions.  All three maps show bimodal distributions spanning the
optically determined cluster centre, although this bimodality is not
seen in redshift space using the limited spectroscopic dataset.  The
presence of substructure 0016+16 and also 1455+22 implies
significant growth in the cluster mass at relatively recent epochs.
This substructure is more readily apparent in the lensing mass maps
than in our X-ray images due to the dependence of the X-ray emissivity
on the smoother cluster potential rather than directly on the mass
distribution.  This graphically illustrates the bias inherent in
determining the prevelance of substructure in clusters on the basis of
X-ray imaging alone. For the intermediate scale substructure crucial to
understanding the inner regions of rich clusters this currently makes
lensing a more sensitive technique.

\noindent{$\bullet$}  Adopting from Paper I the most likely redshift
distribution for the faint galaxy population used as probes in the
lensing analysis we determine the projected mass in the central 1-1.5
Mpc of our two clusters.  These estimates are in extremely good
agreement with those from deprojection of our X-ray images, a result
which is at variance with the conclusions of
Babul \& Miralda-Escud\'e (1994).  The
mass to light ratios obtained in the inner regions of our clusters,
from both the lensing and X-ray analyses, if indicative of the
universal value imply $\Omega \sim 0.8$.

\noindent{$\bullet$} High X-ray luminosity clusters lie on the extreme
tail of the cluster mass distribution in hierarchical models of
structure formation.  The exponential nature of this tail makes the
abundances of such clusters an extremely sensitive test of these
models.  As we demonstrate lensing observations are a direct route to
the underlying mass distribution of these clusters at moderate
redshift, allowing high resolution `imaging' of the cluster mass on
scales comparable to the best available from X-ray imaging.

\noindent{$\bullet$} We have demonstrated the use of weak gravitational
lensing to map the mass distribution in distant clusters.  Enlarging
both the optical and X-ray datasets and the optical area coverage
in individual clusters will allow us to study the prevelance of
mass substructure as a function of epoch in the largest bound
structures known and the role of this substructure in the X-ray evolution
of clusters.  Wider field observations will also allow us to
follow the mass of the clusters out to scales of 2-3 Mpc to determine
the asymptotic form of the mass profile and the run of mass to light in
the outskirts of the cluster (e.g.\ Bonnet, Mellier \& Fort 1994).

\section*{ACKNOWLEDGEMENTS}

We wish to acknowledge the pioneering work of Tony Tyson and the
Toulouse group led by Bernard Fort who first introduced us to the
exciting possibilities gravitational lensing offers in cosmology.  We
also acknowledge useful discussions with Tereasa Brainerd, Carlos
Frenk, Nick Kaiser, Chris Kochanek and Jordi Miralda-Escud\'e.  The
support of the La Palma staff and especially Dave Carter is also
gratefully acknowledged.  This work was supported by the SERC.

\noindent{\bf FIGURES}
\smallskip

\noindent{\bf Figure 1:} a) The ROSAT HRI exposure of 1455+22
overlayed on a grey-scale of our deep combined V+I exposures. b) A
similar comparison between the ROSAT PSPC image of 0016+16
and our V+I exposure.

\noindent{\bf Figure 2:} The upper panel shows
the mass distribution in 1455+22 from the
lensing analysis contoured over the smoothed galaxy number density
distribution for the red cluster members.  The contours start at zero
surface density and are spaced every 2 sigma. The smoothing scale (135
kpc) used to construct both distributions is marked and the position of
the cluster cD is indicated (+).  This map has been normalised by the
local error estimate.  The high degree of similarity between the two
distributions should be noted.  The lower panel
is the surface brightness distribution
of X-ray gas from the ROSAT HRI image of 1455+22 compared to the
derived mass map.  The X-ray distribution's morphology is also very
similar to the mass.  The marked symbols correspond to those in the
previous figure.  The orientation corresponds to that of Figure~1(a).

\noindent{\bf Figure 3:} The four separate tracers available in 1455+22
to map the mass distribution.  Overlayed on these are the best-fit
ellipses to high-light the strong similarities between the orientations
of the distributions over a range of scales.  The ellipse shown for the
X-ray surface brightness map has not been converted into the value for
the mass (c.f.\ Table~1).  Upper-left panel, the lensing derived mass
map.  Upper-right panel, the X-ray surface brightness distribution.
Lower-left, the number density of the red cluster members.  The panel
at lower-right shows the central galaxy.  The orientation and symbols
correspond to those of Figure~2.

\noindent{\bf Figure 4:}
A comparison between the azimuthally averaged profiles for the mass
($\bullet$) and galaxy surface density ($\circ$) in 1455+22.  The
profiles have been corrected for the ellipticity of their respective
distributions.  The more compact nature of the mass distribution is
evident on scales $\gs$ 200 kpc.

\noindent{\bf Figure 5:} This figure shows a more sensitive test of
the relative distribution of galaxies and mass in 1455+22.  By
combining the mass and galaxy surface density maps directly and then
azimuthally averaging we obtain a more direct measure of the
relative concentration of the two distributions.  It is readily
apparent that the mass is more centrally concentrated than the galaxies.

\noindent{\bf Figure 6:}
The total mass, in solar units, interior to 450 kpc from the lensing
analysis of 1455+22 as a function of the median redshift of the
background galaxy distribution ($\bullet$).  This is compared to the
mass derived from the analysis of the X-ray image within the same
radius (solid line).  The predicted value from our preferred redshift
distribution (a no-evolution N(z) to $I=25$, Paper I) is also marked
($\odot$) -- in close agreement to the X-ray measurement.  The redshift
of the cluster is shown by the vertical dashed line.  The total mass
is not corrected for the background surface density at the frame edge.

\noindent{\bf Figure 7:} The left-hand panel
shows the mass distribution in 0016+16 from the
lensing analysis contoured over the smoothed galaxy number density
distribution for the red cluster members.  The contours start at zero
surface density and are spaced every 1 sigma. The smoothing scale (200
kpc) used to construct both distributions is marked and the position of
the optical cluster centre is indicated (+).  This map has been
normalised by the local error estimate.  The right-hand panel
shows the surface brightness
distribution of X-ray gas from the ROSAT PSPC image of 0016+16 compared
to the derived mass map.  The X-ray distribution's morphology is a
close match to the mass.  The marked symbols correspond to those in the
previous figure.   The orientation corresponds to that of Figure~1(b).

\noindent{\bf Figure 8:} Maps for 0016+16 of the four possible tracers
of the cluster mass distribution.  Contoured over these are the
best-fit ellipses showing the strong similarities between their
orientations on scales from 200 kpc to 1.5 Mpc.  The ellipse shown for
the X-ray surface brightness map has not been converted into the value
for the underlying mass.  The lensing derived cluster mass distribution
is shown in the upper-left panel.  The upper-right panel is the X-ray
surface brightness map. Lower-left, the distribution of colour-selected
red cluster members.  The central galaxy distribution is the
lower-right panel.  The orientation and symbols correspond to those of
Figure~7.

\noindent{\bf Figure 9:}
The azimuthally averaged profiles for the mass ($\bullet$) and
galaxy surface density ($\circ$) in 0016+16.  The profiles have been
corrected for the ellipticity of their respective distributions.

\noindent{\bf Figure 10:}
The total projected mass derived from the lensing analysis for the
central 1.2 Mpc of 0016+16.  This is shown in solar units parametrised
in terms of the median redshift of the background galaxy distribution
($\bullet$).  This is compared to the mass determined from the analysis
of the X-ray image within the same radius (solid line).  The predicted
value from our preferred redshift distribution (the no-evolution N(z)
model, Paper I) is also marked ($\odot$) -- in good agreement with the
X-ray measurement.  The redshift of the cluster is shown by the
vertical dashed line.  The projected mass is uncorrected for the
background surface density at the frame edge.


\end{document}